\documentclass[12pt]{article}
\usepackage[margin=1in]{geometry}
\usepackage{setspace}
\usepackage{enumitem}
\usepackage{float}
\usepackage{amsmath}
\usepackage{amsfonts}
\usepackage[none]{hyphenat}
\usepackage{amssymb}
\usepackage{latexsym}
\usepackage{graphicx}
\usepackage[font=small]{caption}
\usepackage[english]{babel}
\usepackage{cite}
\usepackage{tocloft}
\usepackage{tikz}
\usepackage{hyperref}
\hypersetup{
    colorlinks=true,
    linkcolor=blue, 
    citecolor=orange, 
    urlcolor=teal   
}
\addto\captionsenglish{
  \renewcommand{\contentsname}%
    {Contents}%
    \vspace{-1.3em}
}
\begin{document}
\input epsf

\def\p{\partial}
\def\h{{1\over 2}}
\def\be{\begin{equation}}
\def\bea{\begin{eqnarray}}
\def\ee{\end{equation}}
\def\eea{\end{eqnarray}}
\def\d{\partial}
\def\la{\lambda}
\def\eps{\epsilon}
\def\bb{\bigskip}
\def\mm{\medskip}
\newcommand{\dm}{\begin{displaymath}}
\newcommand{\edm}{\end{displaymath}}
\renewcommand{\b}{\tilde{B}}
\newcommand{\gm}{\Gamma}
\newcommand{\ac}[2]{\ensuremath{\{ #1, #2 \}}}
\renewcommand{\ell}{l}
\newcommand{\z}{\ell}
\newcommand{\newsection}[1]{\section{#1} \setcounter{equation}{0}}
\def\bb{$\bullet$}
\def\Qbar{{\bar Q}_1}
\def\QPbar{{\bar Q}_p}

\def\q{\quad}

\def\bn{B_\circ}

\let\a=\alpha \let\b=\beta \let\g=\gamma \let\d=\delta \let\e=\epsilon
\let\c=\chi \let\th=\theta  \let\k=\kappa
\let\l=\lambda \let\m=\mu \let\n=\nu \let\x=\xi \let\r=\rho
\let\s=\sigma \let\t=\tau
\let\vp=\varphi \let\vep=\varepsilon
\let\w=\omega      \let\G=\Gamma \let\D=\Delta \let\Th=\Theta
                     \let\P=\Pi \let\S=\Sigma

\def\h{{1\over 2}}
\def\t{\tilde}
\def\r{\rightarrow}
\def\nn{\nonumber\\}
\let\bm=\bibitem
\def\Kt{{\tilde K}}
\def\b{\bigskip}
\def\m{\medskip}

\let\p=\partial

\newcommand\blfootnote[1]{%
  \begingroup
  \renewcommand\thefootnote{}\footnote{#1}%
  \addtocounter{footnote}{-1}%
  \endgroup
}

\numberwithin{equation}{section}

\newcounter{daggerfootnote}
\newcommand*{\daggerfootnote}[1]{%
    \setcounter{daggerfootnote}{\value{footnote}}%
    \renewcommand*{\thefootnote}{\fnsymbol{footnote}}%
    \footnote[2]{#1}%
    \setcounter{footnote}{\value{daggerfootnote}}%
    \renewcommand*{\thefootnote}{\arabic{footnote}}%
    }

\begin{flushright}
\end{flushright}
\vspace{20mm}
\begin{center}
{\LARGE  The magic of the gravitational vacuum\daggerfootnote{Contribution to `Narlikar's Steady World: Man and the Legend'.}
 }

\vspace{18mm}
{\bf Samir D. Mathur }

\vspace{4mm}

\b

Department of Physics

 The Ohio State University
 
Columbus,
OH 43210, USA

 mathur.16@osu.edu

\b

\vspace{4mm}
\end{center}
\vspace{10mm}
\thispagestyle{empty}
\begin{abstract}

The black hole information paradox challenges us to do something that is seemingly impossible: find a violation of the semiclassical approximation in a region where all curvatures are low. The vecro hypothesis proposes a structure of the gravitational vacuum that can accomplish this task. In this article we explain the hypothesis, and give a lattice model to describe the essence of its idea. The Hamiltonian of the model is completely local, but the vacuum exhibits correlations among planck scale fluctuations which fall off relatively slowly with distance. These extended-scale correlations are able to `feel around' the region where a closed trapped surface is about to form, and to react by nucleating fuzzball structure that  destroys semiclassical spacetime.

\end{abstract}
\vskip 1.0 true in

\newpage
\setcounter{page}{1}


\section{Introduction}

The black hole information paradox \cite{Hawking:1975vcx} has proved deeply puzzling. Many different approaches have been proposed to resolve it. Some have sought to modify the rules of quantum mechanics. Others have abandoned locality, arguing that quantum gravity must contain interactions that do not fall off with distance.

There is, however, good reason to believe that no such drastic measures are needed. In string theory, black holes can be made by taking a bound state at D-branes to sufficiently large coupling \cite{Susskind:1993ws, Sen:1995in, Strominger:1996sh}. The states of these D-branes can be studied in the open string description, where their structure  is governed by an ordinary gauge theory. The gravitational description, in principle, is obtained by rewriting these open string states in terms of closed strings \cite{Maldacena:1997re}. In this latter description we will of course find a complicated structure: the lowest modes of the closed string give the graviton, but there will also be other closed string modes, the various branes of the theory, and intricate bound states of these objects. Nevertheless, this complicated system is still described by a unitary theory. Moreover, the gravitational region describing the black hole will not have nonlocal interactions with distant objects, since the open strings are localized near the branes in the bound state, and the map between open and closed strings is itself local.

In many cases we also know what this complicated region looks like in the gravity description.
 It is possible to explicitly construct many families of black hole microstates, and in each case one finds a {\it fuzzball}; a horizon sized quantum ball, with no horizon \cite{Mathur:1997wb, Lunin:2001jy, Maldacena:2000dr, Skenderis:2008qn, Bena:2015bea, Bena:2016ypk, Jejjala:2005yu, Mathur:2005zp}. This fuzzball radiates from its surface like a normal body, not by pair creation from the vacuum around a horizon.  Thus we evade any  information puzzle. 

But there is one important aspect of fuzzballs that we need to understand: how do fuzzballs form? As we will describe below, semiclassical gravity {\it seems} to be valid in the process of gravitational collapse, which traps the collapsing matter inside a horizon.  If we trust in causality, then no signals can propagate from inside this horizon to the outside. So how does the fuzzball form and send its information out? 

In this article we will explain how this magic can happen, without violating causality or locality at any step. The key postulate is the `vecro hypothesis' \cite{Mathur:2020ely, Mathur:2021zzr, Mathur:2024mtf, Mathur:2025zao}, which says that the gravitational vacuum has a certain set of correlations which are motivated by the fuzzball structure of black holes. We will give a toy example of a vacuum which illustrates the features postulated by the vecro hypothesis, using models studied in condensed matter physics.

\section{The information paradox}\label{sec2}

The strongest form of the information puzzle is as follows:

\b

(a) Consider a shell of mass $M$, composed of massless particles (e.g. gravitons), with each particle moving radially inwards at the speed of light.  In this situation,  no  signal from  any particle can reach  any other particle. Thus each particle should move in a straight line as if the other particles were not present. In particular, each particle should keep moving through the horizon radius of the shell, all the way to $r=0$.

(b) Once the shell has reached a radius $R<2M$, light cones in the region $R<r<2M$ turn `inwards', so that no signals from $r\le R$ can reach the horizon at $r_h=2M$. Thus the region around the horizon remains in the vacuum state, regardless of what new physics might happen at the singularity $r=0$ where the particles meet.

\begin{figure}[ht]
\centerline{
     {\includegraphics[width=3.5in]{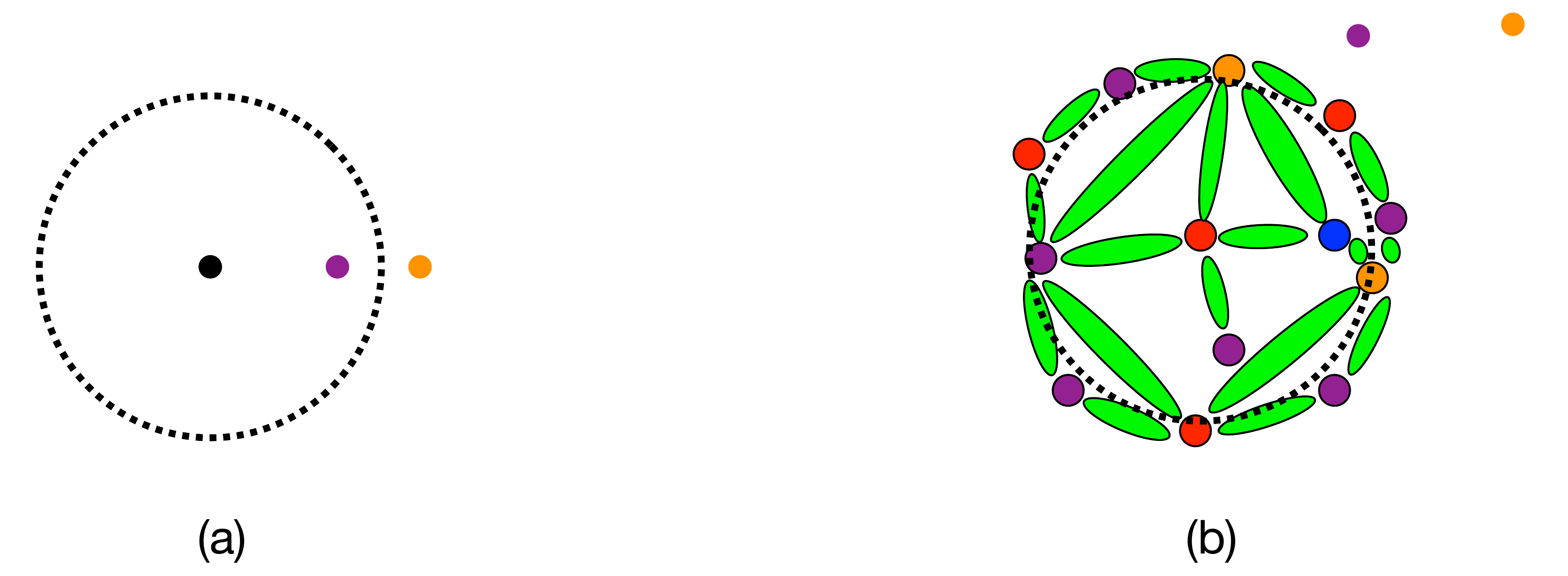}}
}
\caption{(a) The semiclassical hole radiates by pair production from the vacuum around the horizon. (b) The fuzzball radiates from its surface like a normal body.}
\label{fig2}
\end{figure}

(c) By Hawking's computation, particle pairs are produced around the horizon.  One particle escapes to infinity as `Hawking radiation' while the other falls into the hole and lowers its mass (fig.\ref{fig2}(a)).  The two members of the pair are in an entangled state, which may be written schematically as
\be
|\psi\rangle_{pair}={1\over \sqrt{2}} ( |0\rangle_1|0\rangle_2+|1\rangle_1|1\rangle_2)
\label{one}
\ee
Thus the entanglement of the radiation with the remaining hole keeps rising monotonically. If the hole evaporates away completely, then we violate quantum theory. The radiation quanta are entangled, but there is nothing that they are entangled {\it with}; such quanta cannot be described by any quantum state. 

(d) We can try to avoid the problem by saying that the evaporation stops when the black hole reaches planck size. But the resulting `remnant' must have an infinite number of internal states, since we could have started with arbitrarily large $M$ and thus created an arbitrarily large number of entangled pairs.  In string theory,  gauge-gravity duality (i.e., the AdS/CFT correspondence \cite{Maldacena:2000dr}) forbids remnants, since the gauge theory has finitely many states  below any given energy.

(e) One might hope that small quantum gravity effects might  modify Hawking's computation, and somehow `delicately correlate' the emitted quanta so that they  end up in a pure state which is  not entangled with anything. But the small corrections theorem shows this is not possible: if the correction to each pair is $O(\epsilon)$ then the fractional change to the entanglement is also $O(\epsilon)$ \cite{Mathur:2009hf}.   

\b

If a fuzzball forms, then steps (c) onwards are not realized: there is no horizon and so we do not get Hawking's process of pair creation.  But how do we avoid steps (a),(b) and manage to generate a fuzzball instead of the semiclassical black hole? Our goal is to find a structure of the vacuum which will create a fuzzball from the collapsing shell, without violating  locality or causality.

\section{Fuzzballs}

In string theory  a black hole is described by a bound state of the fundamental objects in the theory -- strings and branes. There is considerable evidence that any such bound state swells up into a fuzzball, whose size is always a little larger than the horizon;  this invalidates the semiclassical picture of the black hole \cite{Mathur:1997wb, Lunin:2001jy, Maldacena:2000dr, Skenderis:2008qn, Bena:2015bea, Bena:2016ypk, Jejjala:2005yu, Mathur:2005zp}. The explicit examples of fuzzballs have the following structure. The 9+1 dimensional spacetime has 6 directions compactified to yield a black hole in 3+1 dimensions. Outside the hole, the spacetime is a direct product of 3+1 spacetime with a 6-dimensional compact manifold. But inside the fuzzball, this is no longer the case. A compact circle can shrink to zero size, creating a KK monopole-antimonopole pair. There are topologically  nontrivial $S^2$ surfaces between the centers of such monopoles, and fluxes on these $S^2$s stabilize the entire structure against collapse (fig.\ref{fig2}(b)).

We are not interested here in the detailed structure of fuzzballs. We will instead extract a heuristic picture of the fuzzball that will be adequate for the ideas we wish to present. If we assume that the compact directions are planck scale, then the KK monopoles are objects with planck mass and planck size. Thus we assume that our quantum gravity theory has nonperturbative planck mass excitations, and bound states of these excitations give the microstates of the black hole. 

Note that the fuzzball is an extended object in string theory. This does not mean that the theory is nonlocal in any way. The fuzzball is analogous to a benzene ring in the standard model. The standard model Hamiltonian is completely local, but its fundamental objects -- quarks, gluons, electrons -- make an extended size benzene ring  because this bound state is a local minimum of energy in the theory.

\section{Virtual fluctuations}

The origin of Hawking radiation lies in vacuum fluctuations. We can say that the vacuum is filled with virtual particles, and if the spacetime geometry changes, then these virtual particles can turn into real particles. More precisely, consider a  scalar field $\hat \phi(x)$ satisfying $(\square+m^2)\hat\phi=0$. We can describe this field using a lattice in spacetime. We place a harmonic oscillator at each lattice site, and the gradient term in the Hamiltonian couples the harmonic oscillators at neighboring sites. The Hamiltonian density describing two neighboring sites thus has the schematic form
\be
\hat H =\h  {\hat p^2_1} + \h m \omega^2 \hat\phi_1^2+\h  {\hat p^2_2} + \h m \omega^2 \hat\phi_2^2+k {(\hat \phi_2-\hat \phi_1)^2\over \Delta ^2}
\label{nine}
\ee 
where $\Delta$ is the separation between the lattice sites $x_1, x_2$ where the  two oscillators are based. Since the harmonic oscillators are coupled, the overall ground state is not a product state, but a state of the form
\be
|0\rangle=c_0 |0\rangle_1|0\rangle_2+c_1 |1\rangle_1|1\rangle_2+\dots
\label{ten}
\ee
where $|n\rangle_i$ is the $n$th state of the oscillator at position $i$. 
If the spacetime stretches, then $\Delta$ becomes larger and the coupling between the oscillators at $x_1$ and $ x_2$  becomes smaller. Let us assume that after the expansion of space the oscillators become effectively decoupled, so that excitations of the oscillator $\hat\phi_1$ describe particles at $x_1$ and excitations of the oscillator $\hat\phi_2$ describe particles at $x_2$.  If the expansion is sufficiently rapid, then the wavefunction remains close to (\ref{ten}); this in turn implies that we have `real' particles at $x_1, x_2$ in an entangled state with schematic form (\ref{one}). 

Note that this phenomenon of pair creation  depends crucially on the fact that the state of the field at $x_1$ was correlated with the state at $x_2$ in the form (\ref{ten}).  This correlation arose from the potential given by the last term  in (\ref{nine}). This potential interaction   correlated the fields at $x_1, x_2$ {\it before} the stretching of space; we did not need a light signal to travel between these points in the process of creating the entangled pair.

The above discussion involved a free field $\hat\phi$. But when we consider interactions in the theory, then we find   further correlations among the fluctuations in the vacuum.  The standard model has an interaction between quarks and antiquarks that leads to a bound state -- the meson. This fact leaves its imprint on  vacuum fluctuations. The fluctuations creating a  quark and an antiquark are slightly larger for those configurations where these two particles appear at a separation of about 1 fermi (the meson size); this happens because the energy is lowered for such configurations by the interactions of the theory. Similarly, the vacuum will manifest correlations corresponding to other bound states of the particles in the standard model, like the benzene ring.

Let us now see the effect of such vacuum correlations when the spacetime metric changes. Consider again the shell of radially ingoing quanta considered in section\,\ref{sec2}. When the shell reaches a size smaller than about 3 Angstroms (the radius of the benzene ring), then the part of the vacuum wavefunctional describing the virtual benzene ring will feel an altered gravitational force, and evolve to produce `real' quanta with some small (but nonzero) probability. The crucial point is that these real particles will be produced in a ring around the collapsing shell. It may appear strange that the produced particles have a correlation that reflects the existence of the full collapsing shell, given that the underlying standard model is completely local. But again there is no violation of locality or causality here: the vacuum existed for a long time before the shell arrived, and thus developed nonlocal correlations due to the interactions in the standard model Hamiltonian. The collapsing shell simply converts these correlated fluctuations to  set of real particles, which are correlated to form the shape of a ring.

In string theory we get a similar effect from virtual fluctuations of closed string loops. Suppose the spacetime suddenly starts expanding. Then these virtual loops will get converted, with some probability, to `real' strings.  

While the above effects exist, they are too small to help resolve the information paradox.  They are useful however in illustrating the  path we will take in what follows.  We will look for a structure of the gravitational vacuum that has more correlations than the vacuum in a traditional quantum  field theory.  This conjecture for the gravitational vacuum is called the `vecro hypothesis' \cite{Mathur:2020ely, Mathur:2021zzr, Mathur:2024mtf, Mathur:2025zao}.

\section{The vecro hypothesis}

 It is to be expected that the gravitational vacuum has large fluctuations at the planck scale $l_p$. In the traditional picture of the vacuum, the correlations between these fluctuations are assumed to fall off exponentially for distances $L\gg l_p$.  The vecro hypothesis says that this expectation should be modified: correlations among these planck scale fluctuations should  fall off more slowly; roughly speaking, they should fall off as  a power law. These correlations will react to a collapsing shell and create fuzzball structure, just like in the situation described above where  correlations in the standard model vacuum created real particles in a ring.  
 
We already have a hint for why such extra correlations might exist in the gravitational vacuum. In the standard model, the extra correlations discussed above could be traced to the existence of  bound states in the theory which had an extended size: like a mesons and benzene rings. Are there other bound states that we should think about? Yes, the $Exp[S_{bek}(M)]$ microstates of a black hole, for all values $0<M<\infty$! These bound states are fuzzballs depicted in fig.\ref{fig2}(b), so they have an extended size like the benzene ring, and suggest correlations in the vacuum over extended scales. The vecro hypothesis seeks to model these correlations in a way which will lead to a violation of the semiclassical approximation during gravitational collapse, so that the end result is a fuzzball rather than the traditional hole. The term `vecro' stands for `virtual extended compression resistant objects', and signals that the relevant virtual fluctuations describe extended structures that are resistant to deformation; this resistance leads to the creation of `real' fuzzballs when spacetime is distorted in the process of gravitational collapse. 

\begin{figure}[ht]
\centerline{
     {\includegraphics[width=4.5in]{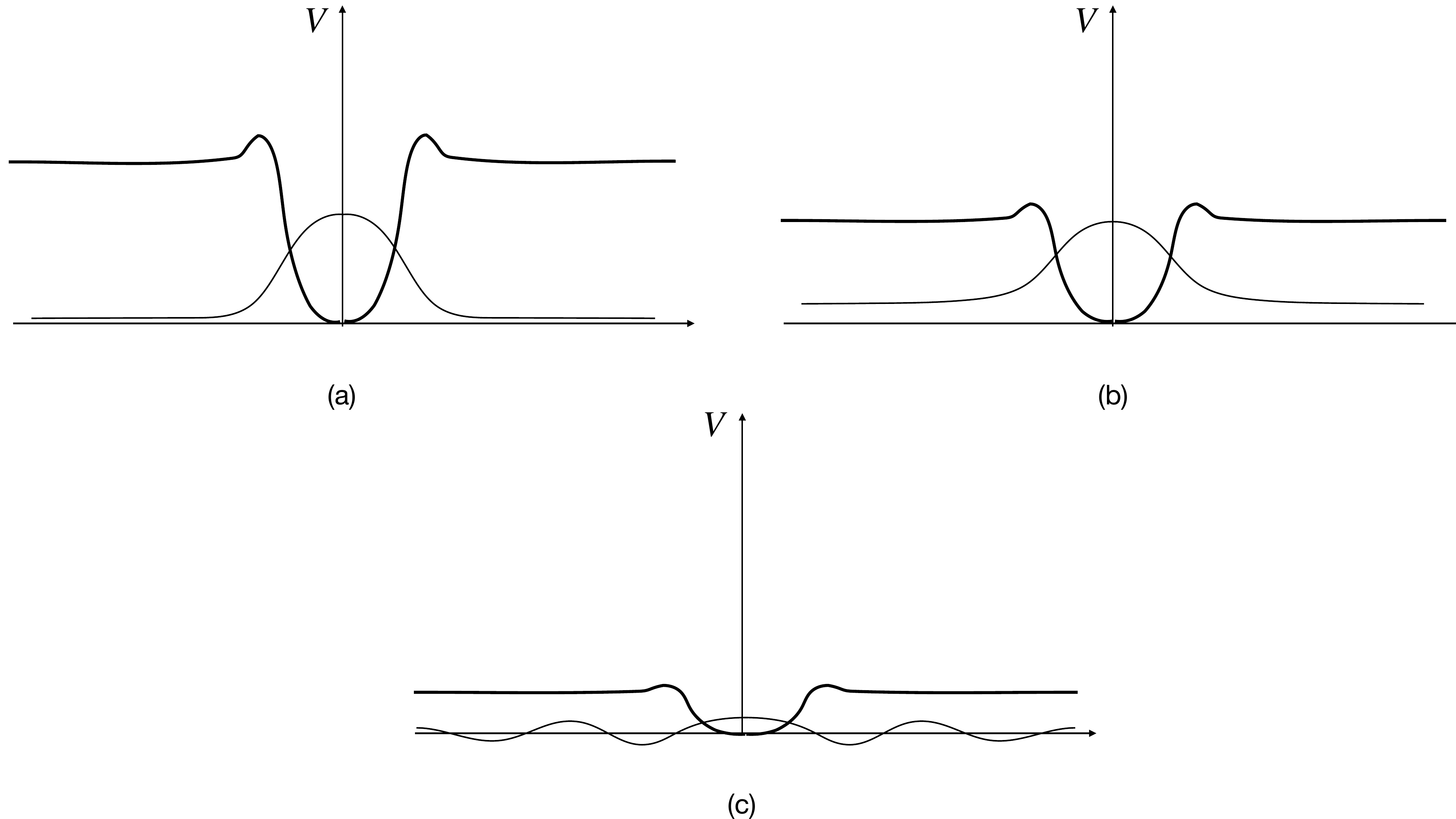}}
}
\caption{(a) The potential (thick line) and the wavefunctional (thin line) for a star. (b) The potential wavefunctional  for a denser star: there is a larger spread over the space of non-semiclassical configurations. (c) Attempting to compress the star inside its Schwarzschild radius leads to a spread over fuzzball configurations.}
\label{fig6}
\end{figure}

Our goal below will be to give a rough understanding of these vecro fluctuations by adapting  models used in condensed mater physics. We will describe the time-independent states of the model, though of course these states can always be superposed  to get dynamical evolution. To relate our discussion of black holes above to this static situation, let us describe how the virtual fluctuations of interest are expected to behave for a {static} star, as we make the star progressively  denser \cite{Mathur:2024ify}:

\b

(a) In fig.\ref{fig6}(a) we depict schematically the energy functional of our quantum gravity theory. On the horizontal axis, we have the various possible configurations of the quantum fields, and the vertical axis is the energy of the configuration. The thick line gives the energy of the configurations of interest. 
The lowest point on the thick line describes a classical star of mass $M$ and radius $R$; thus we have $R\gg 2M$. There are gaussian fluctuations of quantum fields around this classical configuration; these are captured by the quadratic potential around the minimum. But further out, we have a new set of configurations -- the fuzzballs. We focus of fuzzballs of radius $R$, since the wavefunction of the star will have the most overlap with such  configurations But since a fuzzball  has a radius close to its own horizon radius, the fuzzballs of radius $R$ have a mass $M_f=R/2\gg M$. Thus the thick line in fig.\ref{fig6}(a) rises to a high value of the energy $V$ in the region of fuzzball configurations. (The potential `bump' between the gaussian region and the fuzzball region signifies that the fuzzball configurations are accessed by  `tunneling' through intermediate configurations.) 

The thin line in fig.\ref{fig6}(a) depicts the wavefunctional for the star. This wavefunctional is peaked around the classical configuration, but spreads a little around this classical minimum due to the usual gaussian fluctuations of quantum fields. These gaussian fluctuations are responsible for the phenomenon of Hawing radiation. The wavefunctional has a very small tail in the region of fuzzball configurations, since the energy of these configurations is $M_f\gg M$.

\b

(b) In fig.\ref{fig6}(b), the star is denser: its mass is still $M$, but the radius  $R$ is closer to $2M$. Now the fuzzballs with radius $R$ have a mass $M_f$ that is just a little larger than $M$. The wavefunctional in the fuzzball region is now less suppressed: there is still a suppression from the fact that $M_f>M$, but the space of fuzzball configurations is vast (with $Exp[S_{bek}(M_f)]$ configurations), so a nontrivial part of the wavefunctional spreads into  the fuzzball region. This part of the wavefunctional is manifested in the quantum state through virtual fluctuations of bubbles of the kind in fig.\ref{fig2}(b), in the region $r\lesssim  R$. 

\b

(c) In fig.\ref{fig6}(c) we attempt to make a configurations where the star is inside its Schwarzschild radius. Semiclassically, this is of course possible. But in the full quantum gravity theory, the wavefunction no longer `fits' inside the semiclassical part of the potential well; it spreads over the vast space of fuzzball configurations. (If we do try to confine the wavefunction to the gaussian part of the potential, then  the wavefunctional is forced to be `too sharply peaked' so that its energy would be $E>M$; this is  not an allowed state for our object of mass $M$.)

\b

From the above description we can extract the rough features we want from our model of vecro physics.  The Hamiltonian of the theory should be completely local, since we do not find nonlocality in string theory. But the planckian fluctuations of the vacuum should have  correlations that stretch over extended scales. These fluctuations should become larger when a star gets denser, and at the point where semiclassical dynamics predicts a horizon, these fluctuations should dominate the wavefunctional and create a fuzzball instead.

\section{The model}

We are looking for a model where the Hamiltonian has only short distance interactions, but the ground state exhibits correlations over long distances.  We will start with the toric code Hamiltonian \cite{Kitaev:1997wr} and modify it later to suit our goals.\footnote{Toric code models have also been used in the other contexts related to black holes. The topological term in these models gives a subleading correction to area law entanglement \cite{Kitaev:2005dm}. The loops describing the ground state have been related to the loops of loop quantum gravity, and the entanglement generated by such loops has been studied \cite{Feller:2017jqx}. The holographic map in AdS/CFT has been related to quantum error-correction, and the toric code is an example of  such a quantum code \cite{Almheiri:2014lwa, Pastawski:2015qua}. } We will not be interested in the topological properties of this model.  We are also not looking for a model with nonlocal effects across arbitrarily long distance scales; thus we imagine modifying the model later in a way that all effects fall off as a power law with distance.  In what follows, we sketch the essential idea of the model in a simplified way;  a more detailed model will be discussed elsewhere.

\subsection{The basic 2+1 dimensional toric code model}

We begin however by describing the basic toric code model. Consider a square lattice in 2 space dimensions. Place a spin-$\h$ degree of freedom on each edge. For each square plaquette on the lattice, define
\begin{equation}
    B_p = \prod_{i \in \partial p} \sigma_i^z ,
\end{equation}
where the product is over the four edges surrounding the plaquette $p$. Since each $\sigma_i^z$ has eigenvalues $\pm 1$, the plaquette operator also has eigenvalues
    $B_p = \pm 1$.
If an even number of edges around the plaquette have $\sigma_i^z=-1$, then $B_p=+1$. If an odd number have $\sigma_i^z=-1$, then $B_p=-1$.

For each vertex  $s$, which is  typically called the  `star' $s$, define
\begin{equation}
    A_s = \prod_{i \in s} \sigma_i^x ,
    \label{lone}
\end{equation}
where the product is over the four edges meeting at the vertex $s$. Thus $A_s$ flips the four spins adjacent to the vertex $s$. This star operator also has eigenvalues
$    A_s = \pm 1$.

The toric code Hamiltonian is
\begin{equation}
    H=-J_e \sum_s A_s-J_m \sum_p B_p ,
    \label{el}
\end{equation}
where $J_e>0$ and $J_m>0$. The first sum runs over all vertices $s$, and the second sum runs over all plaquettes $p$.

The operators $A_s$ and $B_p$ commute with one another:
\begin{equation}
    [A_s,A_{s'}]=0, ~~~  [B_p,B_{p'}]=0,~~~ [A_s,B_p]=0 \, .
\end{equation}
To see the last relation, note that a star $s$ and a plaquette $p$ either share no edges or share two edges. If they share no edges, the operators trivially commute. If they share two edges then, on each shared edge, $\sigma^x$ anticommutes with $\sigma^z$; thus  $A_s$ and $B_p$ commute.

Since all the terms in the Hamiltonian commute, the energy is minimized by choosing $ A_s=+1$ and $  B_p=+1$ for all $s$ and $p$. An explicit expression for the ground state is
\begin{equation}
    |{\Psi_0}\rangle=\mathcal N \prod_s (1+A_s) |0\rangle
    \label{tw}
 \end{equation}
where $|0\rangle$ is the state with all spins down (i.e., in the state $\sigma^i_z=-1$) and $\mathcal N$ is a normalization constant. One can check that this state satisfies
\begin{equation}
    A_s |{\Psi_0}\rangle = |{\Psi_0}\rangle, ~~~
    B_p |{\Psi_0}\rangle = |{\Psi_0}\rangle
    \label{knine}
\end{equation}
for all $s$ and $p$. To prove the first relation, note that $A_s$ commutes with all factors $(1+A_{s'})$ for $s'\ne s$, while $A_s(1+A_s)=A_s+1$ since $A_s^2=1$. The second relation in (\ref{knine}) is obvious since the $B_p$ commute will all the $A_s$ and $B_p|0\rangle=|0\rangle$.  The ground state energy is
\begin{equation}
   E_0=-J_e N_s-J_m N_p
\end{equation}

\begin{figure}[ht]
\centerline{
     {\includegraphics[width=4.5in]{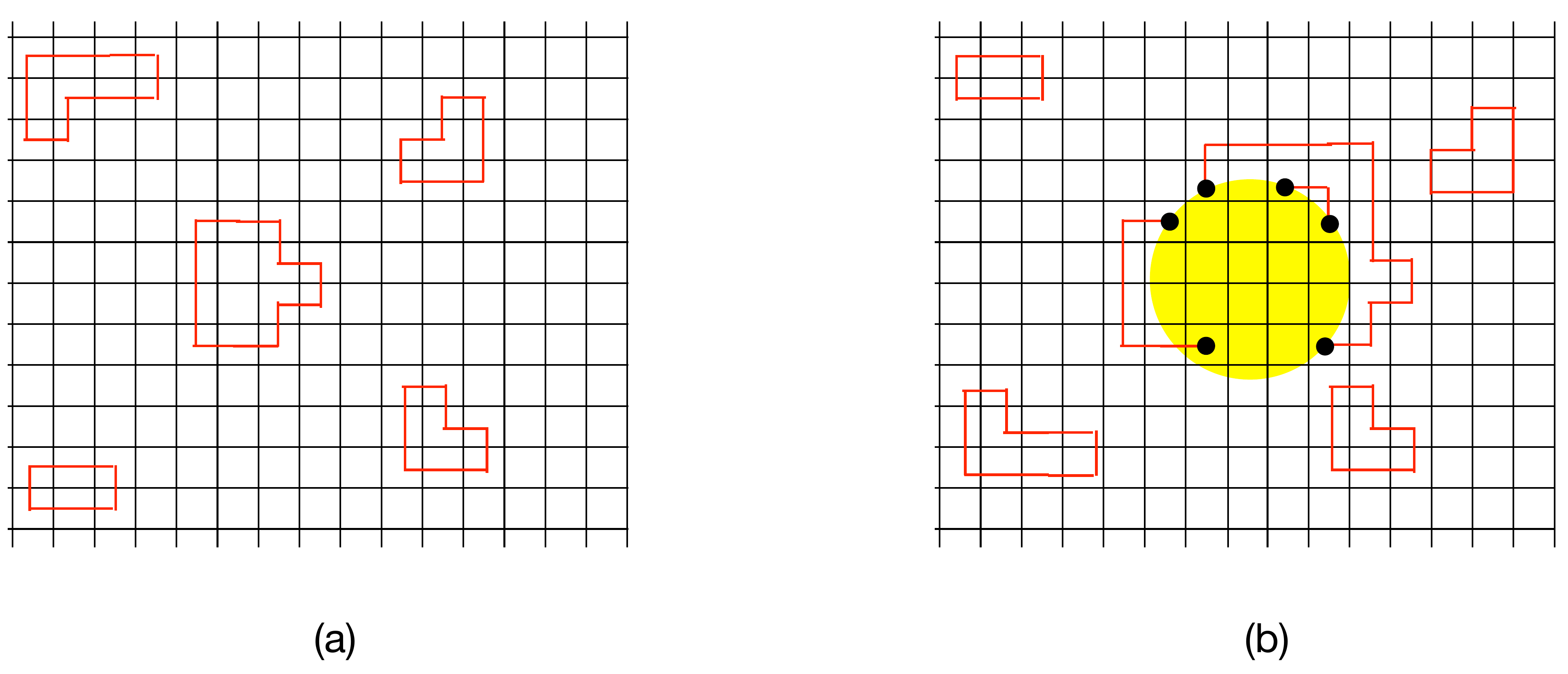}}
}
\caption{(a) The ground state of the toric code is described by a sum over all loops on the dual lattice. (b) The gravitational potential in the central disc leads to loops ending on monopoles; this gives a toy model of fuzzball formation.}
\label{fig1}
\end{figure}

\subsection{Depicting the ground state using paths}

Consider an edge of the lattice. If its spin is in the $s_z=-\h$ state, let us say that the link is `occupied'; if the spin is in the $+\h$ state, let us say that it is `unoccupied'. For an occupied link, we draw a line segment perpendicular to the link; this segment joins the centers of two adjacent plaquettes. 

Now comes the crucial point. In the ground state, we have $B_p=+1$ for every plaquette, so the number of spins around the edges of the plaquette with $s_z=-\h$ is an even number. Thus an even number of the above described segments will meet at the center of each plaquette. We can then join these segments into closed paths on the dual lattice; i.e., the lattice with vertices given by the centers of the plaquettes. If there are $4$ segments meeting at the center of a plaquette, then we can choose a convention of the following type: the bottom and left segments are members of one path, and the top and right segments are members of another path. The ground state (\ref{tw}) can be written as a superposition of terms in each of which each link is in a definite eigenstate of $\sigma^z$. We see that  each term in this sum is described by a set of closed paths $\{ {\mathcal C} \}$ on the dual lattice, as depicted in fig.\ref{fig1}(a). 

Now consider the star terms in (\ref{el}). An operator $A_s$ flips the spins of the $4$ edges that meet at the vertex $s$. On the dual lattice, this  deforms the paths  surrounding the  vertex $s$, in the following way. Consider the $4$ segments of the dual lattice surrounding $s$. If a segment was part of a path ${\mathcal C}$, then it ceases to be a part of ${\mathcal C}$, and if it was not a part of ${\mathcal C}$, then it becomes a part of ${\mathcal C}$.  This effect creates a `gradient' term on the space of paths $\{ {\mathcal C} \}$: the lowest energy state has an equal amplitude for each path ${\mathcal C}$. This can be seen from the explicit expression (\ref{tw}), by expanding the factors and noting the effect of each $A_s$ on the space of paths $\{ {\mathcal C}\}$. 

We will think of these closed paths as the `vecros' of our gravitational vacuum. The spins on our lattice depict planck scale fluctuations of the gravity theory.  The energy is lowered if these fluctuations are correlated with their neighboring fluctuations  in a suitable way; the needed correlations are depicted in our lattice model by the $A_s$ and $B_p$ terms in (\ref{el}). The important point is that the Hamiltonian (\ref{el})  is completely local, but the state of the spins in the ground state is described in terms of closed paths of arbitrary size. Similarly, we can imagine that short distance interactions between  planck scale bubbles in the gravitational vacuum can generate extended correlated structures; these structures are the vecros mentioned above.

\subsection{Electric and magnetic charges}

We saw above that if we wish to minimize the energy of the system, then the condition $B_p=+1$ must hold for each plaquette $p$, and this in turn implies that the number of segments on the dual lattice meeting at the center of $p$ is even. But if we look at excited states, then  we can have $B_p=-1$, which makes the number of segments meeting at the center of $p$ an odd number.  In that case a path ${\mathcal C}$ will have an endpoint at the center of the plaquette $p$, and we say that there is a magnetic charge at $p$. The energy cost of a magnetic charge is $2J_m$.

Similarly, if we have $A_s=-1$ for some vertex $s$, then we say that there is an electric charge at $s$; the energy of such a charge is $2J_e$.

In our gravitational problem, we will identify the magnetic charges with the planck scale bubbles that are depicted by the dots in the fuzzball fig\,\ref{fig2}(b).  Thus a nucleation of magnetic charges in the toric code model will correspond to fuzzball formation in the gravitational theory. 

\subsection{Modifying the model to capture the  `gravitational field'}

The above steps describe the standard toric code model. The Hamiltonian is completely local, but the ground state exhibits a set of closed paths $\{\mathcal C\}$ which extend over all length scales. These paths are analogous to the vecro structures in our gravitational vacuum.

To make a black hole in the gravitational theory,  we need to place matter in some region ${\mathcal S}$, and consider the gravitational field produced by this matter.  We wish to modify the toric code model to include an analogue of this gravitational field. Roughly speaking, the idea will be as follows.  In the toric code model we will say that the paths $\{ {\mathcal C}\}$ cost more energy in a region with gravitational field; thus paths do not want to penetrate into such a region. The operators $A_s$, on the other hand, minimize their energy when the wavefunction {\it does} spread over all paths  $\{ {\mathcal C}\}$. These competing effects lead to a tapering-off of  the path density $P$ in the region with gravitational field. But if this field is increased beyond some strength, then it becomes energetically advantageous for the  paths $\{ {\mathcal C}\}$ to terminate on magnetic charges at the boundary of the region ${\mathcal S}$, rather than for them to continue inside ${\mathcal S}$. This nucleation of magnetic charges will correspond to fuzzball formation. The crucial point is that the vacuum contained extended structures in the form of the paths 
$\{ {\mathcal C}\}$, and these extended structures could `feel around' the entire region ${\mathcal S}$ to detect when a transition to fuzzballs should occur.  But the Hamiltonian was completely local: in particular there are no nonlocal interactions that stretch across the diameter of  ${\mathcal S}$.

In more detail, we proceed in the following steps:

\b

(a) Let the gravitational field at a point $x$ be represented by a function $q(x)>0$. We require that paths at $x$ cost an extra energy, which we write as the `potential energy'
\be
\Delta E_{PE}\sim \int d^{d}x\, q(x) (P(x))^\mu
\label{tone}
\ee
where $P(x)$ is the density of paths at $x$ and $\mu>0$ is a constant that we can choose for the model; for concreteness we set $\mu=2$. In the present case the space dimension is $d=2$. 

Let us see how we could modify the toric code model to include this extra energy (\ref{tone}). There are two ways to think about the gravitational field. We can work with a fixed flat background metric $\eta_{ab}$, and write the field as a correction $h_{ab}$; this approach is suited to the language of a lattice model with a fixed lattice. Alternatively,  we can absorb $h_{ab}$ in the geometry of spacetime and work with the metric $g_{ab}=\eta_{ab}+h_{ab}$. In the latter language the gravitational field has disappeared, but in its place we find that space has been stretched in the gravitating region. Since our lattice structure is fixed, let us use the former description, where the background is a fixed space $\eta_{ab}$ and the gravity encoded in $h_{ab}$ determines $q(x)$. 

The toric code model has a vacuum which is a uniform sum over all closed paths.  To get an effect like (\ref{tone}), we can add a `path tension' which suppresses paths in the region with gravitational field.  This is achieved by adding to the Hamiltonian
\be
H_T={\lambda(x)\over 2} \sum_l (1-\sigma_l^z)
\label{llthree}
\ee
where the sum is over all links  $l$ and $\lambda(x)>0$ in the gravitating region. This Hamiltonian adds an extra energy to spins which are down; i.e., spins which generate a segment of the path ${\mathcal C}$ on the dual lattice. The path tension then suppresses the paths through the gravitating region. 

\b

(b) A star term $A_s$ deforms a path ${\mathcal C}$ to a nearby path ${\mathcal C}'$ which differs from ${\mathcal C}$ by the addition or deletion  of one square of the dual lattice: the square centered at $s$. Thus the star terms in the Hamiltonian (\ref{el}) cause the wavefunction to spread even over all the space of paths $\{ {\mathcal C}\}$; in particular, the ground state of the undeformed model was described by a uniform superposition of all paths ${\mathcal C}$.  We can make a small modification to the toric code model so that the extra `kinetic energy' caused by a nonunifom density of paths is given schematically as
\be
\Delta E_{KE}\sim {1\over G} \int d^d x (\nabla P(x))^2
\label{ttwo}
\ee
Here we assume that $P(x)$ has a value $\bar P$ in flat Minkowski space which is order unity, and  we have used the natural constant $G\sim l_p^{d-1}$ in the problem to set the scale of the energy $\Delta E_{KE}$. 

\b

(c) We see that there is a competition between $\Delta E_{PE}$ and $\Delta_{KE}$. The potential term (\ref{tone}) tries to suppress the path density $P(x)$ in the region with gravitational field, while (\ref{ttwo}) is minimized when the path density is constant. In a general  metric like that of a star, the path distribution takes some form $P(x)$ that minimizes $\Delta E_{PE}+\Delta_{KE}$.  For typical semiclassical configurations, the energy arising from this altered distribution of paths is included in the energy computed by Einstein's action; this description in terms of paths just gives the nature of the wavefunctional that describes the quantum state. But if the gravitational field in increased beyond some point, then it becomes energetically advantageous for the paths to simply end at the boundary of the gravitating region. If we create $N_m$ magnetic charges in this process, then the energy cost of these monopole charges is
\be
\Delta E_m = 2J_e N_m 
\label{tthree}
\ee

\b

(d) If we have
\be
\Delta E_m \lesssim \Delta E_{PE}+\Delta E_{KE}
\label{tfour}
\ee
then the state will transition to one where the paths in the exterior region terminate on a gas of monopoles, instead of continuing inside the gravitating region. We identify this transition with the formation of fuzzballs at the black hole threshold. It was important that the paths -- representing the vecro structures of the gravitational theory -- were extended objects that could feel the whole extent of the gravitating region; a vacuum with such correlations can bypass the naively expected equivalence principle at the boundary of the gravitating region.  

\subsection{The interior of the fuzzball} 

We should also ask what happens to the spacetime {\it inside} the gravitating region. At the boundary of this region ${\mathcal S}$, the path density $P(x)$ has gone to zero -- the paths outside ended in monopoles. Because of the KE term (\ref{ttwo}), the path density cannot suddenly rise inside ${\mathcal S}$ to the  value $\bar P$ which describes the path density  in the vacuum. Suppose the path density does manage to rise to be of order $\bar P$ at the center of gravitating region. Then the kinetic term (\ref{ttwo}) gives an energy
\be
\Delta E_{KE}\sim {1\over l_p^{d-1}} {1\over R^2} R^{d}= {R^{d-2}\over l_p^{d-1}}
\label{tfive}
\ee
where we have noted that $\bar P\sim 1$. This energy cost is not small; in fact it is of the order of the total mass in the gravitating region when we are at the black hole threshold.  To see this, note that the relation $GM_{matter}\sim R^{d-2}$ gives for the mass of the matter at the black hole formation threshold
\be
M_{matter}\sim {R^{d-2}\over l^{d-1}}
\label{tsix}
\ee
Thus (\ref{tfive}) gives $\Delta E_{KE}\sim M_{matter}$. We do not have this extra energy (\ref{tfive}) available in the gravitating region, so we are forced to the configuration where the path density $P(x)$ vanishes throughout the interior of this region. We can interpret this as a breakdown of ordinary spacetime, and its replacement by a fuzzball.

\subsection{Relating the model to the gravitational theory}

The vecro hypothesis says that the Hamiltonian of gravity has only local interactions, but these interactions generate correlations in the vacuum that can be described by extended structures called vecros. In our toy model above we started with the local Hamiltonian of the toric code and noted that the ground state was given by a superposition of  configurations containing paths of all sizes; these paths ${\mathcal C}$ were identified with the vecros of the gravitational theory. We then modified the toric code model so that the path had an additional energy cost when entering a region with gravitational field. For small fields, this resulting path density $P(x)$ and its energy cost are all accounted for by the usual semiclassical gravitational  Lagrangian ${\mathcal R}$; thus the wavefunctional is of the type in fig.\ref{fig6}(a) where semiclassical configurations are the only dominant ones. If we increase the gravitational field strength, then some paths tend to end on monopoles rather than continue into the gravitation field region; this is analogous to fig.\ref{fig6}(b) where a larger part of the wavefunction tunnels into the fuzzball region. Finally, when the gravitational field reaches a certain threshold, there is a percolation type transition and no paths penetrate the gravitational field region; this gives the situation of fig.\ref{fig6}(c) where we have nucleated fuzzballs. 

The important effect in our model came from the density of paths $P(x)$ passing through a point $x$. Since these paths are extended structures, they are able to react to the gravitational field over an extended region.

\section{Breakdown of the semiclassical approximation with ${\mathcal R}\ll l_p^{-2}$}

\begin{figure}[ht]
\centerline{
     {\includegraphics[width=4.5in]{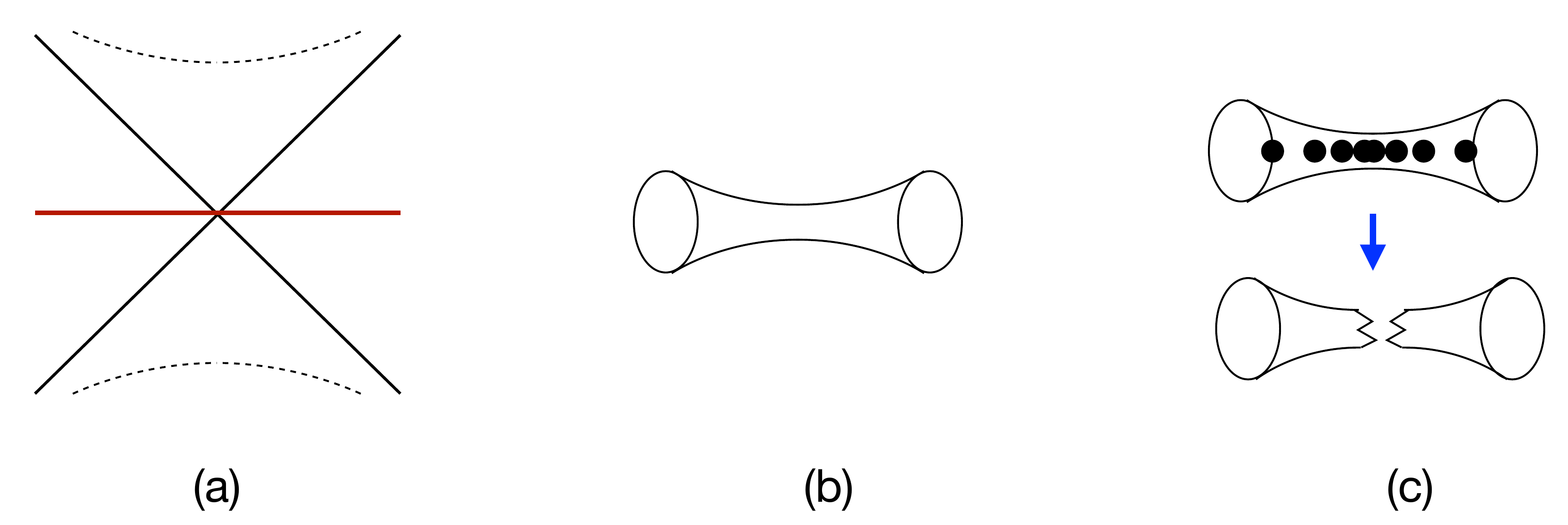}}
}
\caption{(a) The eternal black hole geometry; the central slice is a time-symmetric hypersurface with zero momentum. (b) This slice has the geometry of an Einstein-Rosen bridge, which looks like a smooth initial condition for the semiclassical eternal hole.  (c) In the vecro hypothesis, vacuum bubbles condense near the location of high redshift to minimize the vacuum energy (uuper diagram); this causes the spacetime to degenerate into disconnected fuzzballs (lower diagram).}
\label{fig4}
\end{figure}

The central issue of the black hole information paradox is the following. The semiclassical approximation breaks down when curvatures reach planck scale; i.e., ${\mathcal R} \gtrsim l_p^{-2}$. But is there a second mode of failure of the semiclassical approximation where curvatures are low (${\mathcal R} \ll l_p^{-2}$) but where a closed trapped surface is about to form?

Since our lattice model was time-independent, let us consider a slice of the black hole geometry where the metric is momentarily time-independent. In fig.\ref{fig4}(a) we depict the eternal black hole geometry; the central slice is time-symmetric and thus has zero canonical momenta.   In the geometrical description of gravity, the effect of the gravitational field is encoded in the shape of the slice; this `Einstein-Rosen bridge' is depicted in fig.\ref{fig4}(b). This slice is smooth, with ${\mathcal R} \ll l_p^{-2}$ everywhere. We will now show how the vecro hypothesis disallows this slice as an allowed initial condition for the gravity theory:

\b

(a) The essence of the vecro hypothesis lies in the nature of vacuum fluctuations in the gravitational vacuum. To understand the nature of these fluctuations, let us first recall the fluctuations of a scalar field in the traditional semiclassical black hole geometry.  The eternal black hole resembles 1+1 dimensional Minkowski space in the neighbourhood of the time-symmetric slice. Thus it is tempting to consider the analogue of the  Minkowski vacuum $|0\rangle_M$ on this slice; this vacuum is the Hartle-Hawking vacuum $|0\rangle_{HH}$ which is regular at the horizons. Just like  $|0\rangle_M$ can be decomposed into states of the right and left Rindler wedges, we can write
\be
|0\rangle_{HH}\approx \sum_n e^{-{\beta\over 2} E_n} |E_n\rangle_R|E_n\rangle_L
\label{jone}
\ee
where  $\beta^{-1}$ is the temperature of the  hole and the subscripts $R,L$ label the right and left sides of the eternal hole.

\b

(b) But the state (\ref{jone}) is not the lowest energy state of the scalar field on the eternal black hole geometry. On each of the right and left sides, we can get a lower energy state by taking the analogue of the Rindler vacuum; i.e., the states $|E_0\rangle_R, \, |E_0\rangle_L$.  This suggests that we consider the state $|E_0\rangle_R\,|E_0\rangle_L$ for the scalar field. But we have to be careful about the fact that the Rindler vacuum states are singular near the point where the right and left wedges meet, and the correct state with lowest energy needs regularization at this point. The correctly regulated state has a spike in its energy at this joining point.  We can write the lowest energy state of the scalar field on this slice schematically as
\be
|0\rangle_{lowest}=|E_0\rangle_R\,|E_0\rangle_L~+~ |\psi\rangle_{spike}
\label{jtwo}
\ee
This state is also called the Boulware vacuum. Note that the energy we are talking about here is the energy measured in the Schwarzschild frame and defined using a time direction that moves upwards on each of the right and left sides.  In short, we have to be distinguish the following:

(i) The Hartle-Hawking state $|0\rangle_{HH}$ which is analogous to the Minkowski vacuum $|0\rangle_M$. The Minkowski vacuum is the state with lowest energy when this energy is measured with respect to Minkowski time $T$.

(ii) The state $|0\rangle_{lowest}$ which is the lowest energy state if we define energy with respect to Schwarzschild time (with positive direction upwards on each of the right and left sides).  If we look at the analogue of (\ref{jtwo}) for 1+1 dimensional Minkowski space, then The term $|\psi\rangle_{spike}$ in (\ref{jtwo}) raises the energy of the state  $|0\rangle_{lowest}$ above the energy of $|0\rangle_M$ when energy is measured with respect to the Minkowski time $T$. But when energy is measured using Rindler times, then this spike sits at a point with redshift going to infinity, and the state $|0\rangle_{lowest}$  is the state with lowest energy even after we take into account the energy of the regularized spike.

\b

(c) In the case of a scalar field on the semiclassical eternal black hole geometry, we typically start with the classical geometry of the eternal hole, and then decide which vacuum state of the scalar field we  want to place on this background.  We have a choice of infinitely many vacuum states, of which $|0\rangle_{HH}$ and $|0\rangle_{lowest}$ are two examples. The state $|0\rangle_{HH}$ has a higher energy than $|0\rangle_{lowest}$, but we often consider the state $|0\rangle_{HH}$ since $|0\rangle_{lowest}$ is singular at the point where the right and left wedges join; this singularity creates a singularity along the horizons and makes it unclear how the space should be continued inside these horizons. The reason we have a choice of vacuum state is because the vacuum energy density of the field is considered to be small (i.e.,  higher order in $\hbar$)  than the mass $M$ of the background geometry, and only at locations where the energy density is singular does its backreaction change the leading order classical picture in any significant way. 

\b

(d) But now consider the vacuum state of the gravitational vacuum with the vecro picture of vacuum fluctuations. These fluctuations consist of  planck scale bubbles which must be optimally correlated to reach the state of lowest energy. This optimally correlated state has extended structures called vecros, which are described in our toy model  above by closed paths on the dual lattice. 

 In states where we do not have this optimal correlation of planck bubbles, the energy density will be planck scale, and so the classical geometry will be completely modified. Thus for spacetime described by the Einstein action ${\mathcal R}$ we {\it must} take the planck scale bubbles to be in their lowest energy state. This state is the analogue of $|0\rangle_{lowest}$, since the relevant energy at infinity is the one corresponding to Schwarzschild time. 
 
 But such a state has the spike in energy at the horizon location $r=2M$, similar to  the spike in (\ref{jtwo}). The origin of this spike is the following. There is an increasing redshift as we go towards $r=2M$, and this makes excitations less costly as we approach $r=2M$. Suppose we make a quantum state which makes use of this redshift to lower the overall energy. Then the vacuum excitations closer to $r=2M$ are different from the vacuum excitations further out from $r=2M$. We can see this explicitly through a toy model like (\ref{nine}), where we let the inner oscillator have an energy that is scaled down to a smaller value. The ground state of the coupled oscillators will not be symmetric in the two oscillators; instead there will be larger fluctuations on the inner oscillator which lies at a point of larger redshift. This asymmetry gets more pronounced as we approach $r=2M$ since the gradient of the redshift diverges there. The   vecro structures in the gravitational vacuum must show a singular behavior at $r\r 2M$ with larger and larger fluctuations. In the toy model these fluctuations will consist of monopoles created by the ending of paths. In the gravitational vacuum the fluctuations will describe planck objects of the theory.  Finally, a spike must exist at $r=2M$ since the two sides of the quantum vacuum do not join smoothly at $r=2M$. In the toy model the paths will end at monopoles at this location, and in the gravitational theory the spike describes the fuzzball states.
 
 \b
 
 (e) It is of course known that the Boulware vacuum and its analogues are singular at $r=2M$. But what is crucial in the above discussion is that we {\it cannot} choose the analogue of the Hartle-Hawking vacuum for our planck scale fluctuations. The planck bubbles had a very large energy density, and only the lowest energy configuration of these bubbles will describe a semiclassical spacetime. This lowest energy state must make use of the larger redshift near $r=2M$, and thus necessarily encounter the singular spike describing fuzzballs at $r=2M$. 
 
 It is important to note that the extended nature of the vecro structures was essential to this picture. In the vacuum state for a scalar field there are no such extended structures. Then we can look at the quantum state in an infinitesimal patch around any spacetime point. There is no obstruction to joining these patches together to get the Hartle-Hawking vacuum which is regular on the horizons. But the vecro vacuum has extended structures that relate the region near $r=2M$ with the region farther outside. Then  the overall lowest vacuum state is sensitive to the nature of  the entire slice on which the state is defined. It is crucial that these extended structures are produced by a local Hamiltonian, as in our example of the modified toric code.  There are general theorems that say that for a system with a local Hamiltonian, signals  cannot propagate faster than a given bound. Thus there is no violation of locality or causality at any stage. But the eternal black hole geometry is not allowed as a solution of the gravitational theory.   
 
 \b
 
\begin{figure}[ht]
\centerline{
     {\includegraphics[width=4in]{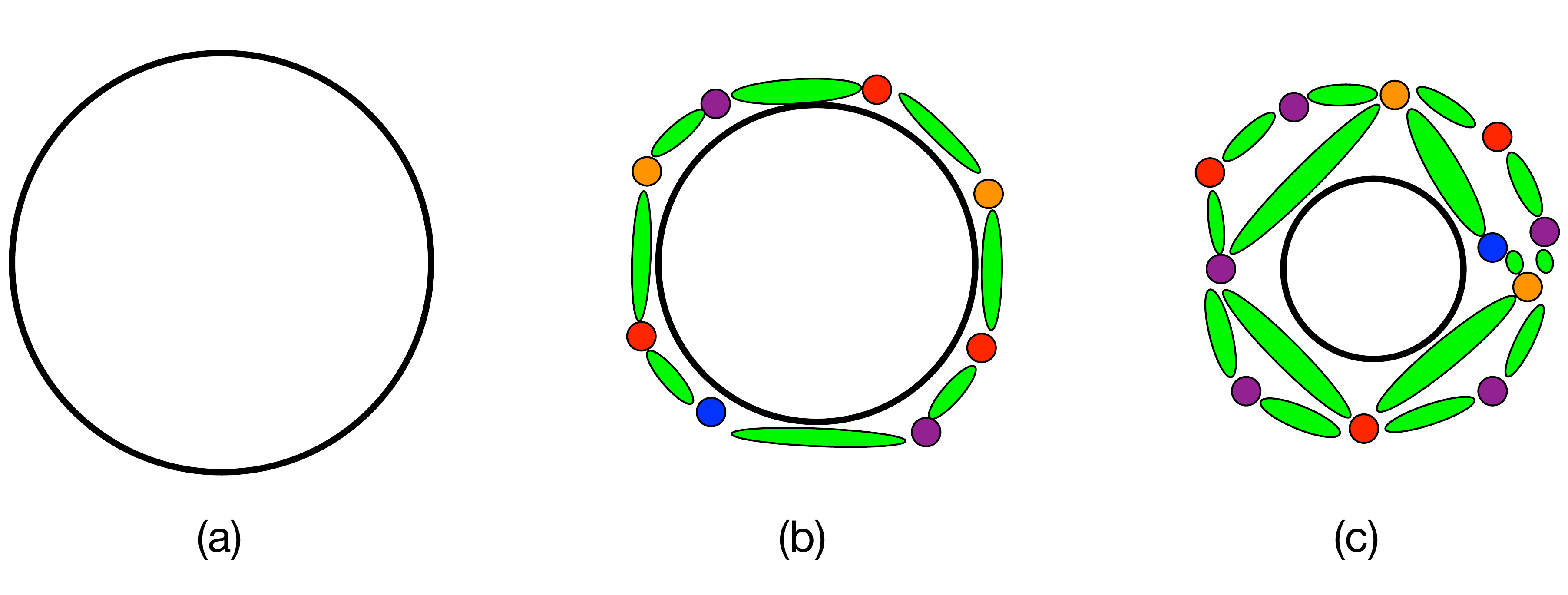}}
}
\caption{(a) A shell collapsing at the speed of light.  (b) Bubbles nucleate as the shell reaches close to its horizon radius, due to the vecro fluctuatins becoming large. (c) These fluctuations join up to create fuzzball structure, resulting in the fuzzball fig.\ref{fig2}(b) as the endpoint of collapse.}
\label{fig3}
\end{figure}

 (f) We discussed the time-symmetric slice of the eternal hole above, but we can also extend the above analysis to the single sided black hole made from collapse, discussed in section\,\ref{sec2}.  Fig.\ref{fig3} depicts the process of fuzzball formation that we expect  in this case. Fuzzballs nucleate outside the shell as the shell reaches its horizon radius, resulting in the fuzzball fig.\ref{fig2}(b) at the endpoint of collapse. 
 
 One interesting aspect of this dynamical problem is the question: what do the quanta on the infalling shell feel during the process of collapse? The conjecture of fuzzball complementarity \cite{Mathur:2015nra} describes how the system could be in the process of nucleating fuzzballs, yet  particles on the shell may feel normal infall (for times of order the crossing time). An analogy can be found in the $c=1$ matrix model, where quanta in the gravitational description emerge as waves on a fermi sea \cite{Das:1990kaa}.  Ripples on the upper edge  of the fermi sea describe ingoing quanta, and ripples on the lower edge of the sea describe outgoing quanta. At the threshold of black hole formation the fermi sea develops `folds' \cite{Polchinski:1994mb}, which destroy the normal map relating the shape of the fermi sea to quanta of the gravity theory \cite{Das:1995gd}. But for short times, we can consider the ripples on a segment at the upper edge of the fermi surface  and map these ripples to ingoing gravity quanta. This toy example illustrates the basic idea of fuzzball complementarity. While the full wavefunctional  is spreading on the space of fuzzballs as in fig.\ref{fig6}(c), we can make an approximate map between this spreading wavefunctional and the  dynamics of infalling quanta. But semiclassical spacetime does not emerge near the horizon: the low energy outgoing quanta which describe Hawking radiation do not emerge in this effective description \cite{Mathur:2015nra}.

\section{Conclusion}

Let us place our considerations above in the context of the information puzzle that we wish to solve. 

Semiclassical dynamics suggests the picture fig.\ref{fig2}(a) for the black hole.  If we accept this picture as a good approximation for low energy dynamics, then we run into trouble: we find the information paradox described in section\,\ref{sec2}.  

In string theory it is possible to explicitly construct the microstates of the hole in many cases. In each case one finds that the microstate is a {\it fuzzball}; a star-like quantum ball with no horizon.  The fuzzball radiates from its surface like a normal body, so there is no information paradox. In \cite{Faraji:2026fdr} it was noted that the classical geometry outside a fuzzball connected naturally to the structure 
fig.\ref{fig2}(b) inside the fuzzball. In \cite{Mathur:2023uoe, Mathur:2024mvo} it was shown that fuzzballs have a universal thermodynamics which matches the thermodynamics obtained from the semiclassical hole. 

Some people thought that the fuzzball resolution of the problem was too radical. They hoped that small corrections to the Hawking process (arising from some hitherto unknown quantum gravity effect) would delicately alter the entanglement among the large number of emitted quanta, in such a way that the radiation ends up in a pure state. But the small corrections theorem shows that such a resolution is not possible. If we assume (a) there are no nonlocal Hamiltonian interactions connecting the region of the hole to the far away region where the radiation collects, and  (b) usual semiclassical dynamics holds to  leading order around the horizon region, then {\it the entanglement of the radiation with the remaining hole will keep rising monotonically} \cite{Mathur:2009hf, Guo:2021blh}. 

String theory has not shown any evidence of the nonlocal effects referred to in assumption (a).  Without such nonlocal effects, the fuzzball resolution of the puzzle is the only option left in the theory. The fuzzball constructions have shown how the various no-hair theorems are bypassed in string theory to yield an object without horizon \cite{Gibbons:2013tqa}.  But we also need to show how the information paradox described in section\,\ref{sec2} is resolved; i.e., we need to show that there is a way to invalidate the semiclassical approximation without having high curvatures in the region of interest or any nonlocality in the underlying Hamiltonian.  The vecro hypothesis gives a model of the quantum gravitational vacuum which achieves this goal.  The model presented in this article explains the essence of the hypothesis: the planck scale fluctuations have correlations that fall off only slowly with distance, and these correlations lead to the emergence of fuzzball structure when enough matter  accumulates in a region to reach the threshold of black hole  formation.

\begin{figure}[ht]
\centerline{
     {\includegraphics[width=4.5in]{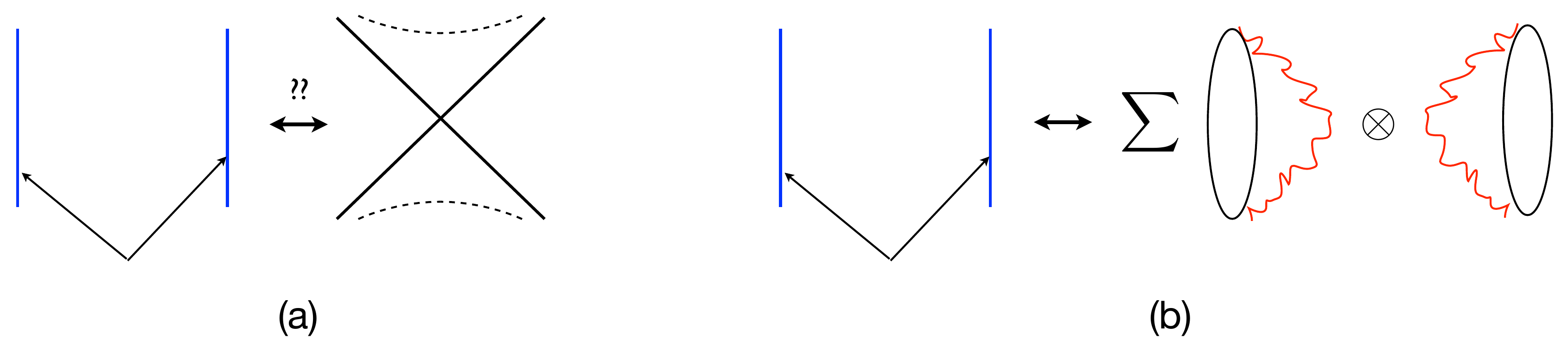}}
}
\caption{(a) The wormhole conjecture says that the dual of two entangled CFTs is geometry with a physical connection between its two sides: the eternal hole. (b) In the fuzzball paradigm, the dual of two entangled CFTs is just a set of entangled fuzzballs, with no Hamiltonian connection between them.}
\label{fig5}
\end{figure}

Finally we would like to contrast the fuzzball paradigm described above with the `wormhole paradigm' which has also been considered in an effort to resolve the information puzzle \cite{Almheiri:2019psf, Almheiri:2019hni, Penington:2019kki, Marolf:2020xie, Liu:2020jsv}. The wormhole idea starts by considering the eternal hole of fig.\ref{fig4}(a), and accepting that this is an allowed solution of the theory. In particular one can consider this eternal hole in AdS.  This geometry has two asymptotic boundaries, so its  dual under the AdS/CFT map  consists of two  CFTs, one at each boundary. These CFTs are noninteracting, though their states can be entangled. This picture gave rise to the notion of ER=EPR: if we entangle the states in two well-separated regions, then we develop a wormhole connecting the two regions (fig.\ref{fig5}(a)). 

But such a notion runs into a problem with the standard rules of string theory. Consider for simplicity type IIA string theory in flat Minkowski space.  Consider a cluster of $N$ D-branes at one position, and a similar cluster a distance $L$ away, with $L\r \infty$. We can entangle the  states  of these two sets of branes, just the way we entangle the spins of two well separated electrons in the lab.  

Each set of branes  can be described by a gauge theory; let these sets  be described by the open string variables $\{ o_1\}, \{ o_2\}$. There is no interaction between the two sets in this description. AdS/CFT duality is just a change of variables from open to closed strings: $\{ o_1\} \leftrightarrow \{ c_1\}, \, \{ o_2\} \leftrightarrow \{ c_2\}$, so the two systems must  still be independent in the closed string (i.e., gravity) description. (There are no open strings stretching between the two systems in the $L\r\infty$ limit, and so we will not get any closed strings connecting the systems either.) If the gravity theory (given by closed string description) somehow  `connected' the two sets of branes, then we could immediately use the above change of variables to write this connection in the language of open strings, i.e., we would get a connection between the two gauge theories. But in standard string theory, two well-separated systems described by gauge theory  do not develop any connection between them. 

This situation forces the advocates of the wormhole paradigm to argue that there must be some nonlocal Hamiltonian effects in string theory, which have not showed up in our calculations with the theory so far. This is an awkward conclusion, so it is worthwhile tracing back to see where one might have gone wrong in the wormhole proposal.  We saw above that accepting the semiclassical picture fig.\ref{fig2}(a) for the `single sided hole' leads to either information loss or remnants. The fuzzball paradigm  resolved this problem by invalidating this semiclassical picture and replacing it by the fuzzball fig.\ref{fig2}(b). Similarly, accepting the semiclassical picture fig.\ref{fig4}(a) for the two-sided hole leads to the above mentioned conflict with the local nature of standard string theory.  In the fuzzball paradigm, we cannot create any geometry with a horizon, so the two-sided hole (which has two sets of horizons) is not an allowed geometry either. In the semiclassical theory, we can imagine starting with the initial slice of fig.\ref{fig4}(b), and allowing it to evolve to create the two-sided hole. But the vecro model described in this article shows how this initial semiclassical geometry is itself ruled out in the theory. With this, the gravitational dual of two entangled CFTs is just two entangled fuzzballs, with no wormhole connecting them (fig.\ref{fig5}(b)). This resolves the conflict with standard string theory mentioned above.

\section*{Acknowledgements}

I am grateful to all the people who have  patiently explained to me  their ideas about the information paradox.  I am grateful to Madhur Mehta and Emil Martinec for discussions. This work  is supported in part by DOE grant DE-SC0011726.


\end{document}